\documentclass[pra,amsmath,amssymb,twocolumn]{revtex4}

\usepackage[dvips]{graphicx}

\usepackage{epsfig}

\begin{document}

\title{Calculation of quantum discord in higher dimensions for $X$- and other specialized states}

\author{A. R. P. Rau\footnote{A. R. P. Rau \\
Department of Physics and Astronomy, Louisiana State University, Baton Rouge, Louisiana 70803, USA \\
email: arau@phys.lsu.edu}}
\affiliation{Department of Physics and Astronomy, Louisiana State University, Baton Rouge, Louisiana 70803, USA}

\begin{abstract}
{\bf Abstract}   Quantum discord, a kind of quantum correlation based on entropic measures, is defined as the difference between quantum mutual information and classical correlation in a bipartite system. Procedures are available for analytical calculation of discord when one of the parties is a qubit with dimension two and measurements made on it to get that one-way discord. We extend now to systems when both parties are of larger dimension, of interest to qudit-quDit with $d, D \geq 3$ or spin chains of spins $\geq 1$. While recognizing that no universal scheme is feasible, applicable to all density matrices, nevertheless a procedure similar to that for $d=2$ that works for many mixed-state density matrices remains of interest as shown by recent such applications. We focus on this method that uses unitary operations to describe measurements, reducing them to a compact form so as to minimize the number of variables needed for extremizing the classical correlation, often the most difficult part of the discord calculation. Results are boiled down to a simple recipe for that extremization; for some classes of density matrices, the procedure even gives trivially the final value of the classical correlation without that extremization. A qutrit-qutrit ($d=D=3$) system is discussed in detail with specific applications to density matrices for whom other calculations involved difficult numerics. Special attention is given to the so-called $X$-states and Werner and isotropic states when the calculations become particularly simple. An appendix discusses an independent but related question of the systematics of $X$-states of arbitrary dimension. It forms a second, separate, part of this paper, extending our previous group-theoretic considerations of systematics for qubits now to higher $d$. 


\end{abstract}

\maketitle

\section*{1 Introduction}

Quantum correlations other than entanglement \cite{ref1} have increasingly been discussed in quantum information. Among them, quantum discord has attracted attention and is calculated in terms of von Neumann entropies of density matrices \cite{ref2}. Thereby, its definition and calculation is in principle available for bipartite systems AB of arbitrary dimension. This contrasts with entanglement, where for mixed states, convenient and reliable measures such as concurrence or negativity have not been established beyond qubit-qubit and qubit-qutrit (dimension three) systems in general \cite{ref3} except for partial results in continuous systems \cite{ref4}. The problem is that once the dimension of both A and B equals or exceeds three, whether a measure such as negativity \cite{ref5} is zero or non-zero does not uniquely distinguish between separable and non-separable states as it does for qubit-qubit and qubit-qutrit. Indeed, this is a NP hard problem \cite{ref6}, as is also the general calculation of quantum discord \cite{ref7}, also clear from the fact that identities exist that relate the two correlations \cite{ref8}.  

Nonetheless, the motivation for studying quantum correlations such as discord in higher dimensions is not only the use of qutrits and higher qudits in quantum information but the study of spin chains with spin larger than 1/2. Thus spin chains of spin 1, with Heisenberg interaction among spins and single ion anisotropies, have been studied for strongly correlated magnetic systems through numerical diagonalization and density matrix renormalization group techniques \cite{ref9}. A richer variety of phenomena than exhibited by spins 1/2 are of interest, including quantum phase transitions and their transition points, richer phase diagrams, etc. \cite{ref9}. Other studies of higher spins and higher dimensional systems \cite{ref10} and experimental advances \cite{ref11} also add to the interest in quantum discord. 

And, the continuing appearance in the literature of discord calculations for certain sub-classes of states of higher dimension have motivated the work we present here. Based on the key idea, first given in \cite{ref12} for a qubit, of conditional measurements as a combination of projectors and unitary transformations, that was developed into a convenient scheme in \cite{ref13,ref14}, we extend that procedure now to higher dimensional qudits. Within the limitations on general applicability set by these being NP hard problems, we address the efficient handling of the unitary transformations in $d$ dimensions for calculating classical correlation and thereby discord for some classes of mixed states. While we consider von Neumann projectors, we will also comment on the more general POVM (positive operator valued measures).

A convenient procedure for describing and computing the measurements involved in discord calculations is to parametrize unitary transformations of a qubit as given in \cite{ref12,ref13}. An overstatement in \cite{ref13} was corrected later \cite{ref14} along with simultaneously reducing the discord calculation to a very simple prescription applicable to qubit-qudit or $N$-qubit systems. The largest number of variables over which the classical correlation has to be maximized is two, although simplifications often reduce the work involved even further. In extending now to cases when both parties are of dimension larger than that of a qubit, and in particular for a qutrit or spin 1 with dimension three, the largest number of variables needed is six, becoming even fewer for classes of density matrices of most interest. We consider $X$-states \cite{ref15,ref16} as well as Bell, Werner, and many other states that are often discussed in a variety of situations. Although not restricted to or directly connected to discord, we provide in an appendix a systematic definition of such $X$-states, their symmetry groups and number, in any dimension, again for current and future interest.

The arrangement of this paper is as follows. Section 2 reviews the procedure for calculating discord when a qubit is involved \cite{ref14}. This presentation is focussed on one of our primary concerns, to reduce to a minimum the number of parameters that need to be varied for maximizing the classical correlation, so as to point to the same for higher dimensions. Many-parameter variation can be tedious so that reducing the number of them is very important. Section 3 presents such a procedure and illustrates with the example of a qutrit of dimension 3. Section 4 applies the method to compute quantum discord for some qutrit-qutrit density matrices, of both pure and mixed states, of interest. An alternative geometric discord has been presented for some of these \cite{ref17} and we use that and other work for comparison. We discuss especially $X$-states where, for arbitrary dimension, the form of the density matrix is preserved, thus making calculations simpler; at the same time, these states encompass much physics of interest. 

The appendix presents a group symmetry of these states that makes transparent their generation and properties, again connecting to and generalizing what is available for qubits \cite{ref14,ref16,ref18}. These results in the appendix may be of interest more generally than for the particular correlation of discord, since they express general properties and symmetries of qudit-quDit systems.

\section*{2 Procedure for a qubit} 

A bipartite system AB with A a qubit and B of possibly larger dimension such as four with a density matrix  

\begin{equation}
\rho =  \left( 
\begin{array}{cccc|cccc}
\rho_{11} & 0 & 0 & \rho_{14} & \rho_{15} & 0 & 0 & \rho_{18} \\ 
0 & \rho_{22} & \rho_{23} & 0 & 0 & \rho_{26} & \rho_{27} & 0 \\ 
0 & \rho_{32} & \rho_{33} & 0 & 0 & \rho_{36} & \rho_{37} & 0 \\
\rho_{41} & 0 & 0 & \rho_{44} & \rho_{45} & 0 & 0 & \rho_{48} \\
\hline
\rho_{51} & 0 & 0 & \rho_{54} & \rho_{55} & 0 & 0 & \rho_{58} \\
0 & \rho_{62} & \rho_{63} & 0 & 0 & \rho_{66} & \rho_{67} & 0 \\
0 & \rho_{72} & \rho_{73} & 0 & 0 & \rho_{76} & \rho_{77} & 0 \\
\rho_{81} & 0 & 0 & \rho_{84} & \rho_{85} & 0 & 0 & \rho_{88} 
\end{array}
\right)      
\label{eqn1}
\end{equation}
serves to illustrate our previous results \cite{ref13,ref14} applicable to B of any dimension but with A assumed to be a qubit. Such a so-called ``extended-$X$" state \cite{ref14} can be viewed as four equally-sized blocks, such a $2 \times 2$ block structure reflecting sub-system A; within each block, the $4 \times 4$  denotes the dimension of sub-system B, these blocks having the structure of the letter X with non-zero entries only along the diagonal and anti-diagonal \cite{ref15,ref16}. The extended-$X$ states are a subset of the general density matrix with non-zero entries everywhere in Eq.~(\ref{eqn1}). 

In calculating quantum discord, the quantum mutual information in the full system AB, defined as \cite{ref2,ref12}  

\begin{eqnarray}
\mathcal{I} (\rho^{AB}) = S (\rho^A) + S (\rho^B) - S(\rho^{AB})\, , \label{eqn2}
\end{eqnarray}
where $S(\rho) = - \mathrm{tr} \, ( \rho \, \log_2 \rho )$ is the von Neumann entropy, is easily computed, even analytically. Note that in doing so, for the eigenvalues required for these entropies, the 1-4-5-8 and 2-3-6-7 subspaces of Eq.~(\ref{eqn1}) are decoupled, simplifying the algebra involved.  

\subsection*{ 2.1  Computing classical correlation}

The first part being the quantum mutual information, the second part of calculating discord is the computing of classical correlation between A and B by accounting for all possible projective measurements over A. The essential idea \cite{ref14} is to use von Neumann projections, as transformed by arbitrary unitary transformations, to describe all possible measurements on qubit A. Although for higher dimensions, the more general POVM (positive operator valued measure) operators are needed, for qubits, it is known that the von Neumann projectors are enough to describe the most general measurement. Stated physically, any general spin-1/2 measurement means the two antipodes on a sphere in a Stern-Gerlach arrangement so that the simple projectors suffice along with covering all points on the sphere. Starting with two orthogonal projectors, $\Pi_i = |i\rangle\langle i |, \, i=\pm \hat{z} $ for sub-system A, and unitary operators $U \in SU(2)$, a general measurement for A can be written as \cite{ref12}
\begin{eqnarray}
A_i = U \, \Pi_i \, U^{\dagger} . \label{eqn3}
\end{eqnarray}
$U$ may be parametrized in terms of Pauli spin matrices of A as \cite{ref12}

\begin{eqnarray}
U = t \, I + \mathrm{i} \, \vec{y}\cdot\vec{\sigma} \, ,\label{eqn4}
\end{eqnarray}
with $t,y_1,y_2,y_3 \in \mathbb{R}$ and $t^2 + y_1^2 + y_2^2 + y_3^2 = 1$. 
Only three of these parameters are independent, assuming values $t ,  y_i \in [-1,1]$ for $i = 1,2,3$.

The conditional density operator $\rho_i$ associated with a measurement of sub-system A is \cite{ref12}
\begin{eqnarray}
\rho_i = \frac{1}{p_i} (A_i \otimes I) \rho (A_i \otimes I) \,,
\label{eqn5}
\end{eqnarray}
where the probability $p_i$ equals $\mathrm{tr} [(A_i \otimes I) \rho (A_i \otimes I)]$. Because of the projection operators in it, we have $A_i ^2 =A_i, \sum_i A_i =I$. With the choice of measurement directions $i = \pm \hat{z} $, we have

\begin{equation}
A_{\pm}= U  \Pi_{\pm z}  U^{\dagger} = (I \pm \vec{\sigma} \cdot \vec{z})/2.
\label{eqn6}
\end{equation}

The seemingly complicated calculation in Eq.~(\ref{eqn5}) with $A_i$ on either side of the density matrix reduces to a very simple prescription at the end. A crucial identity for this purpose, which follows from Pauli matrix identities, is that for any vector $\vec V$,

\begin{equation}
A_{\pm} (\vec{\sigma} \cdot  \vec {V}) A_{\pm} = \pm (\vec z \cdot \vec V) A_{\pm} .
\label{eqn7}
\end{equation}
Here, $\vec{z}$ is a unit vector formed out of the four parameters in Eq.~(\ref{eqn4}) as given in \cite{ref14}

\begin{equation}
\vec{z} = \! \{2(-t y_2 + y_1 y_3),  2(t y_1 + y_2 y_3), t^2 + y_3^2 - \! y_1^2 -\! y_2^2 \}. 
\label{eqn8}
\end{equation}  

The above procedure, first given in \cite{ref12}, already contains a subtlety at this point to which we will return in Sec. 3, namely, a reduction from the three parameters in Eq.~(\ref{eqn4}) to just the two of the unit vector $\vec{z}$ in Eq.~(\ref{eqn8}). An alternative standard parametrization of the unit vector in polar angles, the usual `Bloch angles' \cite{ref1}, gives the $2 \times 2$ matrix in Eq.~(\ref{eqn6}) as \cite{ref14}

\begin{equation}
A_{+}= \left(
\begin{array}{cc}
\cos ^2 (\theta/2) & \frac{1}{2}\sin \theta \exp (-i\phi)  \\
\frac{1}{2}\sin \theta \exp (i\phi) & \sin^2 (\theta/2) 
\end{array}
\right),
\label{eqn9}
\end{equation}
and $A_-$ its parity conjugate with $(\theta,\phi)$ replaced by $(\pi -\theta,\pi +\phi)$, that is, with diagonal entries interchanged and a change in sign of the off-diagonal entries.

The conditional density matrix \cite{ref12} for sub-system B because of measurements on A is given by 
\begin{eqnarray}
S (\rho | \{A\}) = \sum_i p_i \, S(\rho_i) \, ,
\label{eqn10}
\end{eqnarray}
and its quantum mutual information, a measure of total correlation \cite{ref19}, by

\begin{eqnarray}
\mathcal{I} (\rho|\{A\}) = S (\rho^B) - S(\rho|\{A\}) \, . \label{eqn11} 
\end{eqnarray}
A measure of the resulting classical correlations then follows \cite{ref2,ref12} as
 
\begin{eqnarray}
\mathcal{C}(\rho) = \sup_{\{A\}} \, \mathcal{I} (\rho|\{A\}) \, . \label{eqn12} 
\end{eqnarray}
Finally, the quantum discord is obtained as the difference between Eq.~(\ref{eqn2}) and Eq.~(\ref{eqn12}),

\begin{eqnarray}
\mathcal{Q}(\rho) = \mathcal{I}(\rho) - \mathcal{C}(\rho) \, .
\label{eqn13}
\end{eqnarray}

\subsection*{2.2  Qudit discord recipe}

As we have shown before \cite{ref14}, the entire calculation in Sec. 2.1 can be condensed into the form of a recipe or a simple prescription for any such system AB with A a qubit. All that is needed is to take elements of each $\{jk \}$-block of the four blocks of Eq.~(\ref{eqn1}), multiply by the $kj$-th element of Eq.~(\ref{eqn9}), and add the four blocks together to get $p_i \rho_i$ in Eq.~(\ref{eqn5}) after which eigenvalues and entropies in Eq.~(\ref{eqn10}) and Eq.~(\ref{eqn11}) are easily calculated. The supremum in Eq.~(\ref{eqn12}) requires varying the two angle parameters in Eq.~(\ref{eqn9}). Because of the parity symmetry that has been noted \cite{ref13}, the expression is always an even function of $\cos \theta$. 

Further, because of the way $\phi$ occurs in Eq.~(\ref{eqn9}), for many density matrices, this parameter disappears from the eigenvalue calculation and the extremization reduces to just one variable $\theta$. Even further, although our original conclusion \cite{ref13} that the extremum is reached at the value of $\theta = \pi/2$ is only true when the function has only one extremum and an actual computation is necessary in general to determine the value of $\theta$, that conclusion (overstatement) seems to hold for all but a tiny fraction of density matrices \cite{,ref14,ref20,ref21}, thereby making the calculation even simpler in most instances.

In the spirit of providing in the form of a recipe, we note that although it is an obvious consequence of labelling rows and columns of entries and blocks in Eq.~(\ref{eqn1}) from left to right in AB, when it is B that is the qubit and A of possibly any dimension, the equivalent recipe for discord calculation with measurements over the qubit end is to take each of the $2 \times 2$ blocks in the qudit-qubit $\rho$, multiply the $jk$-th element by the $kj$-th element of $A_i$ in Eq.~(\ref{eqn9}) and add all four to give the $d \times d$ conditional density matrix. This same prescription applies to the more general considerations in Sec. 3, blocks added together for measurements made on the left A end but all within a block added when the measured end is B on the right. Of course, for symmetric systems such as those considered in Sec. 4, the entropic discord either way is the same. 

\section*{3 Discord calculation for spin larger than 1/2}

The above procedure when A is a qubit is, in principle, easily extended to when it (and B) is of larger dimension, say $d_A \geq 3$. The density matrix of AB would, as in Eq.~(\ref{eqn1}), be viewed in terms of $d_A \times d_A$ blocks, each block a matrix of $d_B \times d_B$. With $d_A$ projectors $\Pi_i$ in Eq.~(\ref{eqn3}), a natural parametrization of the unitary matrix in that equation would involve $(d_A ^2 -1)$ independent operators and coefficients in Eq.~(\ref{eqn4}). Thus, for A a spin-1 or qutrit with $d_A=3$, eight Gell-Mann matrices $\lambda_i$ \cite{ref22} would occur instead of the Pauli matrices. Two of them are diagonal and also would stand in the projectors:

\begin{equation}
\Pi_{1,2} =(2I \pm 3\lambda_3 +\sqrt{3}\lambda_8)/6, \, \, \Pi_3 =(I -\sqrt{3}\lambda_8)/3.
\label{eqn14}
\end{equation}
A calculation analogous to that in Sec. 2 would give $d_A$ matrices A as in Eq.~(\ref{eqn6}) and Eq.~(\ref{eqn9}). Multiplying their $kj$-th element into the $jk$-th block of $\rho_{AB}$, and adding the blocks would give the $d_B \times d_B$ conditional density matrix from which eigenvalues and classical correlation are computed.

However, an immediate question arises, namely, what is the number of independent parameters needed for the extremization? We saw in the previous section that for a qubit it is at most two, not the expected 3 of SU(2) unitary matrices. Understanding this reduction provides the clue to the generalization to the higher SU($d_A$). For this purpose, our previous casting of a unitary $U$ for any SU($N$) as a product $U_1U_2$ with the first term describing a ``base manifold" and the second a ``fiber" is useful \cite{ref23} and we give a brief summary in this paragraph. For the SU(2) of a qubit, these matrices are exponentials of the Pauli matrices, with the fiber $U_2$ diagonal, being the exponential of $\sigma_z$. Since it commutes with the projectors in Eq.~(\ref{eqn6}) that also involve $\sigma_z$ only, it disappears at this stage of the calculation, leaving the A matrices in that equation dependent only on $U_1$ and, therefore, on that base manifold's two dimensions only. Instead of the Cartesian Pauli matrices, it is convenient \cite{ref24} to consider $\sigma_{\pm}$ when $U(1)$ is a product of two matrices with unity along the diagonal and one non-zero off-diagonal entry of a complex number $z$ that is given by a simple first-order differential equation. Through an inverse stereographic projection, that $z$ becomes the unit vector of the Bloch sphere with its polar angles as in Eq.~(\ref{eqn9}). (There is here an interesting parallel, two parameters appearing either as the complex number $z$ of \cite{ref23} or as the unit vector $\vec{z}$ of Eq.~(\ref{eqn8}).) This picture generalizes to higher dimensions in a very natural fashion in Secs. 3.1 and 3.2 below. 

This picture of unitary operations in terms of a base manifold and a fiber, the latter not occurring in the extremization and the former simply related to the Bloch sphere, gives a ready explanation for the reduction to two variational parameters and a natural geometrical casting of them. It is the key to our simplification compared to other results on qutrit systems in the recent literature \cite{ref25,ref26,ref27}. Therefore, we will now extend this argument to qutrits and higher-dimensional systems, a principal goal of this paper being to provide an explicit route to calculating the minimal set of parameters needed for computing quantum discord.     

\subsection*{3.1 Example of a qutrit}

In repeating the calculation of classical correlation and thereby quantum discord, we use for concreteness the value $d_A=3$ of a qutrit to illustrate the procedure that, however, applies more generally. For the $3 \times 3$ case, the first question is how many parameters will be involved in the extremization in analogy to the previous section's reduction from three to two. That it will again not take the full eight of SU(3) symmetry, that is, $(d_A^2-1)$ real parameters, as may be expected at first sight, is known, the number of real parameters characterizing one-dimensional orthogonal projectors for von Neumann measurements being $d_A(d_A-1)$. This makes for 6 for a qutrit. For a convenient and explicit construction for qutrits of results analogous to Eq.~(\ref{eqn6}) and Eq.~(\ref{eqn7}), we use as stated in the previous paragraph a similar decomposition of $U$ as $U_1U_2$ that has been provided in \cite{ref24}. 

The $2 \times 2$ structure of the qubit's SU(2) is retained by regarding the $ 3 \times 3$ SU(3) matrices blocked so as to be again $2 \times 2$ but now in blocks, the upper diagonal block itself a $2 \times 2$ matrix.  Although $U_1$ in Eq.(28) of that paper is a linear combination of all eight $\lambda$, it depends only on four parameters, viewed either as two complex numbers $(z_1,z_2)$ or four real angles $(\theta_1, \theta_2, \epsilon_1, \epsilon_2)$. The second unitary matrix $U_2$, the fiber, is now block-diagonal with an upper $2 \times 2$ SU(2) block and a lower $1 \times 1$ single entry. It is clearly a linear combination of just four of the Gell-Mann matrices $(\lambda_{1-3}, \lambda_8)$. Therefore, in the analog to Eq.~(\ref{eqn6}), although the full $U_2$ no longer commutes through to disappear in a product expression, there is nevertheless considerable simplification. 

For this purpose, the structure of $\Pi_i$ in Eq.~(\ref{eqn3}) is crucial, that they are matrices with only one non-zero entry, namely unity in the $i$-th diagonal position, $i=1,2, \ldots d_A$. The view of $U$ in Eq.~(\ref{eqn3}) in terms of $U_1U_2$ as in \cite{ref23,ref24}, with two unitary factors, the latter block-diagonal in terms of lower dimensional $U$ means that the only parameters needed are the $z$ of \cite{ref23,ref24} involved at each step in such a construction of $U$. These sets of complex numbers $z$, or their alternative rendering in terms of pairs of polar angles, are the parameters varied to extremize the classical correlation. 

For A a qutrit, $U_1$ is given in Eq. (27) and Eq. (28) of \cite{ref24}. $U_2$ has an upper diagonal block of a qubit and a pure phase as its lower diagonal block. As a result, for $A_3$ in Eq.~(\ref{eqn3}), with $\Pi_3$ having only one non-zero entry of unity in the lowest diagonal position, $U_2$ drops out to leave behind $U_1 \Pi_3 U_1^{\dagger}$ which further becomes

\begin{equation}
A_3 \!\!=\!\! \left(
\begin{array}{ccc}
s_1^2 c_2^2 & s_1^2s_2c_2 e^{i(\epsilon_1 -\epsilon_2)} & -s_1c_1c_2 e^{i \epsilon_1} \\
s_1^2s_2c_2 e^{-i(\epsilon_1 -\epsilon_2)}& s_1^2 s_2^2 & -s_1c_1s_2 e^{i \epsilon_2} \\
-s_1c_1c_2 e^{-i \epsilon_1} & -s_1c_1s_2 e^{-i \epsilon_2} & c_1^2
\end{array}
\right),
\label{eqn15}
\end{equation}
where we we have abbreviated $ c= \cos \theta, s=\sin \theta$ with subscripts 1 and 2 for the $(\theta_1, \theta_2)$ in the previous paragraph.

Thus, $A_3$ involves only four parameters, those of the two complex $z$ of the SU(3) base manifold in Eq. (27) and Eq. (28) of \cite{ref24}. $A_1$ and $A_2$ are a little more complicated expressions than Eq.~(\ref{eqn15}), involving both the two complex $z$ of the SU(3) $U_1$ but also the two parameters in Eq.~(\ref{eqn6}) and Eq.~(\ref{eqn8}) of their SU(2) complex $z$ in $U_2$, again using the abbreviations $c$ and $s$ for the trigonometric functions in Eq.~(\ref{eqn9}). They are evaluated by sandwiching between $U_1$ and $U_1^{\dagger}$ the $3 \times 3$ block-diagonal matrix that has in its upper diagonal block $A_{\pm}$ from Eq.~(\ref{eqn6}) and unity in the last diagonal position. They depend, therefore, on 6 parameters, as also noted in \cite{ref9}. Thus, we have

\begin{eqnarray}
(A_1)_{11} & = & c^2(s_2^2+c_2^2c_1)^2 + s^2c_2^2s_2^2(1-c_1)^2 \nonumber \\
 &  & - 2csc_2s_2(1-c_1)(s_2^2+c_2^2c_1) \cos (\epsilon_1\!\!-\!\!\epsilon_2\!\!+\!\!\phi), \nonumber \\
(A_1)_{12} & = & cs[(s_2^2+c_2^2c_1)(c_2^2+s_2^2c_1)e^{-i\phi} \!\!+s_2^2c_2^2(1-c_1)^2 \nonumber \\
 &  & \times e^{i(2\epsilon_1-2\epsilon_2+\phi)}]-f e^{i(\epsilon_1-\epsilon_2)}, \nonumber \\
(A_1)_{13} & = & css_1s_2[(s_2^2\!\!+\!\!c_2^2c_1)e^{i(\epsilon_2-\phi)}\!\!-\!\!c_2^2(1-c_1)e^{i(2\epsilon_1\!\!-\!\!\epsilon_2\!\!+\!\!\phi)}] \nonumber \\
 &  & +[c^2(s_2^2+c_2^2c_1)-s^2s_2^2(1-c_1)]s_1c_2e^{i\epsilon_1}, \nonumber \\
(A_1)_{22} & = & c^2s_2^2c_2^2(1-c_1)^2+s^2(c_2^2+s_2^2c_1)^2 \nonumber \\
 &  & - 2css_2c_2(c_2^2+s_2^2c_1)(1-c_1) \cos (\epsilon_1\!\!-\!\!\epsilon_2\!\!+\!\!\phi), \nonumber \\
(A_1)_{23} & = & css_1c_2[(c_2^2+s_2^2c_1)e^{i(\epsilon_1+\phi)}-s_2^2(1-c_1) \nonumber \\
 &  & \times e^{-i(\epsilon_1-2\epsilon_2+\phi)}]- g e^{i\epsilon_2}, \nonumber \\
(A_1)_{33} & = & c^2s_1^2c_2^2 \!\!+\!\!s^2s_1^2s_2^2 \!\!+\!\! 2css_1^2s_2c_2 \cos (\epsilon_1\!\!-\!\!\epsilon_2 \!\!+\!\!\phi), 
\label{eqn16}
\end{eqnarray}
with $A_2$ a similar expression in which $c^2$ and $s^2$ are interchanged and the sign of $cs$ is changed. We have defined for convenience: $f =[c^2(s_2^2+c_2^2c_1)+s^2(c_2^2+s_2^2c_1)]c_2s_2(1-c_1)$ and 
$g = [c^2c_2^2(1-c_1)-s^2(c_2^2+s_2^2c_1)]s_1s_2$. All three matrices $A_i$ are Hermitian and unitary with unit trace and we have $A_1+A_2+A_3=I$. These properties, together with the feature noted after Eq.~(\ref{eqn5}) that $A_i^2=A_i$, will prove crucial in applications in Sec. 4. 

The calculation of quantum discord for a qutrit system therefore requires at most the 6 parameters, angles $(\theta, \theta_1,\theta_2)$ and phases $(\epsilon_1, \epsilon_2, \phi)$ in Eq.~(\ref{eqn15}) and Eq.~(\ref{eqn16}), not the original 8 of the qutrit's su(3) algebra. Further, we will see in Sec. 4 that for many density matrices, not even all of this smaller set are needed (just as for qubits, often one parameter extremization suffices \cite{ref14}); in particular, most of the phase parameters drop out in computing eigenvalues leaving primarily the $(\theta, \theta_1, \theta_2)$ parameters in the extremization to get the classical correlation and, thereby, the quantum discord. This cuts the number of parameters by half.

\subsection*{3.2 Qudits of arbitrary dimension}

Extension to dimensions larger than a qutrit proceeds similarly in our construction of $U$ for a general SU($N$). For A a qu4it with $d_A=4$, the density matrix $\rho_{AB}$ is viewed as a $4 \times 4$ block matrix, each block of $d_B \times d_B$. In constructing $A_i$ for the $4 \times 4$ SU(4) matrix, $U$ is again regarded as a product $U_1U_2$ of three matrices, all decomposed into blocks either as $4=2+2$ or $4=3+1$. According to the procedure in \cite{ref23}, in the former decomposition, $U_1$ which itself is a product of two matrices, will involve diagonal blocks of the unit matrix and one non-zero off-diagonal $2 \times 2$ matrix of 4 complex numbers $z$. The matrix $U_2$ has two diagonal blocks of $2 \times 2$ matrices, and these SU(2) are each regarded as in Sec. 2 with its own complex number $z$. In all, the $A_i$  has a total of 6 complex $z$ or 12 real parameters. The alternative block reduction of $4=3+1$, followed by $3=2+1$, will mean again 3+2+1 or 6 complex $z$ in the reduction of SU(4) in stages through SU(3) and SU(2). 

The case of general $d_A$ proceeds similarly, involving the sequence of triangular numbers for the number of complex $z$ needed, that is, $d_A(d_A-1)$ real parameters in all, just as expected. Instead of the canonical representation in terms of Gell-Mann matrices or their higher-dimensional counterparts, the view of \cite{ref23,ref24} through a sequence of $z$ (or, alternatively, a polar and an azimuthal/phase angle) provides the desired reduction in the number of parameters needed for extremization and an explicit step-by-step realization of them. It is, therefore, the preferred route to calculating quantum discord or in possible other applications. This is where we differ from previous attempts at discord calculations for qutrit-qutrit that also handled through Gell-Mann matrices \cite{ref25,ref26,ref27}.

\section*{ 4   Applications to qutrit-qutrit systems}

To illustrate the general procedure of Sec. 3, we consider some density matrices of qutrit-qutrit systems that have been recently studied in different contexts and for whom the geometric discord, a correlation simpler to evaluate, have been computed \cite{ref17} as well as entropic discord \cite{ref28} and other correlations \cite{ref25,ref26,ref27}. 

\subsection*{4.1 Bell states}

Labelling as usual the three states as 0, 1, and 2, we begin with the maximally entangled Bell state, a pure state described by $(|00\rangle+|11\rangle+|22\rangle)/\sqrt{3}$ and density matrix of the form

\begin{equation}
\rho =  \left( 
\begin{array}{ccc|ccc|ccc}
\rho_{11} & 0 & 0 & 0 & \rho_{15} & 0 & 0 & 0 & \rho_{19} \\ 
0 & 0 & 0 & 0 & 0 & 0 & 0 & 0 & 0 \\ 
0 & 0 & 0 & 0 & 0 & 0 & 0 & 0 & 0 \\
\hline
0 & 0 & 0 & 0 & 0 & 0 & 0 & 0 & 0 \\
\rho_{51} & 0 & 0 & 0 & \rho_{55} & 0 & 0 & 0 & \rho_{59} \\
0 & 0 & 0 & 0 & 0 & 0 & 0 & 0 & 0 \\
\hline
0 & 0 & 0 & 0 & 0 & 0 & 0 & 0 & 0 \\
0 & 0 & 0 & 0 & 0 & 0 & 0 & 0 & 0 \\
\rho_{91} & 0 & 0 & 0 & \rho_{95} & 0 & 0 & 0 & \rho_{99} 
\end{array}
\right),      
\label{eqn17}
\end{equation}
all non-zero entries equal to 1/3 for the Bell state. By inspection, eight of the eigenvalues are zero and one unity so that the last term in Eq.~(\ref{eqn2}) is zero, all as befitting a pure state. Tracing over A or B in Eq.~(\ref{eqn17}) gives $3 \times 3$ density matrices $I/3$ with corresponding entropies in Eq.~(\ref{eqn2}) of $\log 3$. Conventionally in quantum information, logarithms have been evaluated in base 2 but it seems to fit better for our discussion of qutrits to use base 3 (more generally, base $d$) so that the total mutual information in Eq.~(\ref{eqn2}) equals 2. This also fits what follows in the next paragraph for the classical correlation.

The procedure of Sec. 3 for evaluating the reduced density matrix and its entropy becomes trivial. With all blocks in Eq.~(\ref{eqn17}) having one non-zero entry of 1/3, the prescription of multiplying by Eq.~(\ref{eqn15}) or Eq.~(\ref{eqn16}) and adding blocks gives back those same $A_i$ matrices, multiplied by 1/3. Given that these matrices are Hermitian and unitary, the eigenvalues are 0, 0, and 1 so that the entropy in Eq.~(\ref{eqn10}) of these reduced $3 \times 3$ matrices is zero. Hence, no supremization is necessary and, from Eq.~(\ref{eqn11}) and Eq.~(\ref{eqn12}), the classical correlation equals $\log 3 =1$, again using base 3. So, just as with Bell states of qubit-qubit, we can conclude that all such maximally entangled Bell states of qutrit-qutrit, or any qudit-qudit, have $\mathcal{I} =2$, equally divided into classical correlation and quantum discord of $\mathcal{C} =\mathcal{Q} =1$. 

\subsection*{4.2 Werner, isotropic, and pseudo-pure states}

The Bell states of the previous sub-section are pure states. A class of mixed states obtained by mixing them with a completely mixed state that is essentially the unit matrix, $I/d$, with equal diagonal elements and all off-diagonal coherence terms zero (``white noise"), has been much studied both for qubit-qubit and higher dimensions. Terminology (and notation) has varied in referring to them as Werner  \cite{ref17,ref25,ref26,ref27}, isotropic, or ``pseudo-pure (PP)" states \cite{ref28}. Consider the isotropic state given by the density matrix,

\begin{equation}
\rho_{AB} = [(1-p)/d^2]I + p|\Psi\rangle\langle \Psi|,
\label{eqn18}
\end{equation}
with $0 \leq p \leq 1$ and $|\Psi\rangle$ the Bell state as in Eq.~(\ref{eqn17}) with all non-zero entries equal to $1/d$. Since the discussion for general $d$ is no more complicated than for $d=3$, we will consider such states of any dimension.

The evaluation of quantum discord is straightforward. For a qudit-qudit state in Eq.~(\ref{eqn18}), by inspection, one eigenvalue is $p+(1-p)/d^2$ and the remaining $d^2-1$ eigenvalues are $(1-p)/d^2$, and $S_{AB}$ in Eq.~(\ref{eqn2}) follows immediately. Tracing over either one of the qudits clearly gives a completely mixed state, $I/d$, so that $S_A = S_B = \log d$ in Eq.~(\ref{eqn2}). Next, the calculation of the conditional density matrix in Eq.~(\ref{eqn5}) is equally straightforward. Given the linearity of that expression, each term in Eq.~(\ref{eqn18}) can be treated separately. The first, proportional to the unit operator, commutes through the two $A_i$ and, with the square being $A_i$ with unit trace, is proportional to a $d$-dimensional unit matrix. And, as noted in Sec. 4.1, in our procedure for a Bell state, the second term also gives a reduced $d \times d$ density matrix that is $A_i$ to a multiplicative factor. With our observation in Sec. 3 that the eigenvalues of $A_i$ are all zero except one which equals unity, the eigenvalues of the conditional density matrix thereby reduce to $d-1$ of magnitude $(1-p)/d$ and one with $[p+(1-p)/d]$. The results of Eq.~(\ref{eqn11}) and Eq.~(\ref{eqn12}) are then immediate, making unnecessary any supremum calculation because all the parameters in $A_i$ have dropped out. This seems to have escaped notice in \cite{ref17,ref26,ref27}, the ``quite difficult" supremization carried out although all angles involved took zero values in the final results. This was in itself a pointer to the parameters introduced by $A_i$, or other SU(3) counterparts in their calculations, dropping out of the calculation. 

The classical correlation in Eq.~(\ref{eqn12}) and quantum discord in Eq.~(\ref{eqn13}) are simple algebraic expressions evaluated from these eigenvalues,

\begin{equation}
\mathcal{C}= [(d\!-\!1)(1\!-\!p)/d] \log(1\!-\!p) +[p+(1\!-\!p)/d] \log(dp+1\!-\!p),
\label{eqn19}
\end{equation}
and

\begin{eqnarray}
\mathcal{Q} &\!\! =\!\! & [(d\!-\!1)(1\!-\!p)/d^2] \log(1\!-\!p) +[(d^2p \!+\!1\!\!-p)/d^2]  \nonumber \\
 \!\!&  &\!\! \times \log (d^2p\!+\!1\!\!-\!\!p) \!\! -\!\![(dp\!+\!1\!-\!p)/d] \log (dp\!+\!1\!\!-\!\!p).
\label{eqn20}
\end{eqnarray}
These results, valid for all $d$, coincide for $d=3$ with the expressions given as Eqs.(9) and (10) in \cite{ref26} for qutrits. Therefore, instead of repeating the plots in that reference of these correlations as a function of $p$, we tabulate instead as an alternative presentation our values (using base 3 for logarithms) in Table I. 

Limiting results for these correlations are interesting. For $d$ large, and assuming that $p$ is neither too close to 0 or 1 so that $pd >> 1-p$, we have $ \mathcal{Q} = p \log d $ rising linearly with $p$, a result also in \cite{ref28}. With the choice of base $d$ for any dimension, this reduces to the mixing parameter $p$, independent of dimension. This might be used as another justification for using logarithms to the base $d$, the observation \cite{ref28} of correlations growing without bound with $d$ being merely a reflection of the growth in $\log_2 d$ when the base is fixed at 2. We also note that \cite{ref27} generalized the results of \cite{ref26} to what are called ``Werner derivatives" by introducing another parameter besides $p$. Again, a laborious supremization was carried out but is unnecessary as per our procedure, the final correlations simple algebraic expressions depending on the two parameters for any $d$. The maximization of classical correlations seems to not enter into such states, correlations independent of measurements on one of the sub-systems, no doubt because of the mixing of isotropic configurations in the density matrix. 

\begin{table}[h] 
\begin{center}
\begin{tabular}{|c|c|c|c|c|}
\hline
 $p$&$S_{AB}$&$\mathcal{C}$&$\mathcal{Q}$&$\mathcal{I}$ \\ 
 \hline
$0$&$2$&$0$&$0$&$0$   \\
\hline
$0.1$&$1.97$&$0.008$&$0.022$&$0.03$   \\
\hline
$0.2$&$1.89$&$0.034$&$0.072$&$0.106$ \\
\hline
$\frac{1}{4}$&$1.84$&$0.053$&$0.106$&$0.159$ \\
\hline
$\frac{1}{3}$&$1.74$&$0.094$&$0.169$&$0.263$ \\
\hline
$0.4$&$1.64$&$0.135$&$0.226$&$0.361$ \\
\hline
$\frac{1}{2}$&$1.47$&$0.21$&$0.324$&$0.534$ \\
\hline
$\frac{2}{3}$&$1.11$&$0.377$&$0.509$&$0.886$ \\
\hline
$\frac{3}{4}$&$0.903$&$0.485$&$0.612$&$1.097$ \\
\hline
$0.9$&$0.441$&$0.735$&$0.824$&$1.36$ \\
\hline
$0.95$&$0.25$&$0.85$&$0.90$&$1.75$ \\
\hline
$1$&$0$&$1$&$1$&2 \\
\hline
\end{tabular}
\end{center}
\caption{Correlations for the qutrit Werner state in Eq.~(\ref{eqn18}) as a function of the mixing parameter $p$. The entropies $S_A$ and $S_B$ equal 1, while $S_{AB}$, classical correlation, quantum discord, and total mutual information are shown.}
\end{table}

\subsection*{4.3  $X$-states}

There is an extensive qubit literature for what are called $X$-states. Originally so named \cite{ref15} because of the appearance of a $4 \times4$ qubit-qubit density matrix with non-zero entries only along the diagonal and anti-diagonal resembling the letter X, their attraction lies in the fewer parameters involved in them. With trace fixed, 3 real elements along the diagonal and 2 complex ones in the off-diagonal amount to 7 parameters in all. As has been noted \cite{ref29}, local unitaries can render the off-diagonals also as real numbers to make for 5 parameters in all. This is less than the 15 of a general qubit-qubit density matrix and makes for easier handling. At the same time, many specialized states of interest belong to this class and many physical phenomena may be studied with them without necessarily involving more general matrices. 

For bipartite systems with only one a qubit, the slightly larger class of ``extended $X$" states as in Eq.~(\ref{eqn1}) also prove almost as convenient because the conditional density matrix also ends up as an $X$-state. Eigenvalues are then evaluated through simple quadratic equations, calculations breaking into $2 \times 2$ sub-spaces. For qubit-qubit, extended $X$-states include all possible $4 \times 4$ density matrices with 15 parameters \cite{ref14} of such a system. Through various local transformations, higher dimensional density matrices of a more general nature can also be brought into $X$-form \cite{ref30,ref31}, lending further importance to them. (A different direction of generalization \cite{ref31} to what have been termed ``true generalized $X$" (TGX) states gives states that differ from our extended $X$ states.)

It seems then natural to look at similar $X$-states in dealing with higher-dimensional systems. Thus, a qutrit-qutrit state with entries only along the diagonal and anti-diagonal,

\begin{equation}
\rho^{(X)} =  \left( 
\begin{array}{ccc|ccc|ccc}
\rho_{11} & 0 & 0 & 0 & 0 & 0 & 0 & 0 & \rho_{19} \\ 
0 & \rho_{22} & 0 & 0 & 0 & 0 & 0 & \rho_{28} & 0 \\ 
0 & 0 & \rho_{33} & 0 & 0 & 0 &  \rho_{37} & 0 & 0 \\
\hline
0 & 0 & 0 & \rho_{44} & 0 & \rho_{46} & 0 & 0 & 0 \\
0 & 0 & 0 & 0 & \rho_{55} & 0 & 0 & 0 & 0  \\
0 & 0 & 0 & \rho_{64} & 0  & \rho_{66} & 0 & 0 & 0 \\
\hline
0 & 0 & \rho_{73} & 0 & 0 & 0 & \rho_{77} & 0 & 0 \\
0 & \rho_{82} & 0 & 0 & 0 & 0 & 0 & \rho_{88} & 0 \\
\rho_{91} & 0 & 0 & 0 & 0 & 0 & 0 & 0 & \rho_{99} 
\end{array}
\right)      
\label{eqn21}
\end{equation}
has 16 parameters, fewer than the 80 of a completely filled $9 \times 9$ matrix. Again, this results in a great economy in handling so that these states are natural targets for studying quantum correlations and other characteristics of qudit-quDit systems. In the Appendix, we will consider a symmetry structure of these states just as has been described for qubits \cite{ref16,ref18} that also permits extension to multipartite systems.

For a discord calculation with such $X$-states, eigenvalues involved in the calculation of Eq.~(\ref{eqn2}) are simple, as it is in the case of qubits, the central 5-5 element itself an eigenvalue and the rest as pairs of $2 \times 2$ sub-spaces. The same applies to the reduced densities upon tracing over either A or B, each a $3 \times 3$ matrix with once again the central element decoupled from the others. For the conditional density upon measurements over A, our prescription in Sec. 3 of multiplying each of the 9 blocks by the corresponding transposed element of $A_i$ and adding the blocks clearly results in an $X$-state again. Eigenvalues again follow easily. Also, as can be seen from Eq.~(\ref{eqn15}) and Eq.~(\ref{eqn16}) for a qutrit, parameters in their 1-2 and 2-3 elements drop out, reducing the number required for supremization. 

Whereas our prescription applies quite generally to any qutrit-quDit density matrix giving a $D \times D$ conditional density, for $X$-states and extended $X$-states, there is such a reduction from 4 to 3 ($\epsilon_2$ dropping out) for $A_3$ and from 6 to 5 for $A_{1,2}$. In particular, many of the phase angles drop out leaving a small subset of them plus the theta angles which are $d_A(d_A-1)/2$ in number. Mutual phases between successive sub-spaces of SU($d_A$) involved in the reduction of unitary matrices, as per \cite{ref23}, add $d_A-2$ for a total of approximately $d_A(d_A+1)/2$ parameters to be varied, which is also the number of POVMs (Positive Operator Valued Measures \cite{ref32}). As noted in Sec. 2, POVMs are not necessary, von Neumann projectors sufficient in the case of qubits where these numbers are 3 and 2, respectively but this is not rigorously true for qutrits. Nevertheless, our Sec. 3 sticks with the three orthogonal projectors, without invoking six POVMs, because of generating classical measurements through the use of a general $U$ along with the $\Pi_i$ in Eq.~(\ref{eqn3}). Indeed, in employing such a combination of $U$ and von Neumann projectors, we also have finally six parameters for qutrits. For larger dimensions, our similar combination's number of parameters $d_A(d_A-1)$ actually exceeds the number of POVMs.  

Finally, we note also other studies with qutrits \cite{ref33,ref34,ref35} where our prescription could be applied to simplify the numerics of the extremization involved in them. The form of the density matrix considered in \cite{ref33} is interesting. This qutrit-qutrit density matrix differs from the $9 \times 9$ matrix in Eq.~(\ref{eqn21}) in having X character for the five blocks along diagonal and anti-diagonal but an ``anti-X" structure for the other four $3 \times 3$ blocks, that is, zeroes now along the diagonal and anti-diagonal within the blocks. As a result, application of our prescription in Sec. 3 of multiplying the blocks by elements from Eq.~(\ref{eqn15}) and Eq.~(\ref{eqn16}) and then adding leads to a conditional density matrix with non-zero elements for all nine entries. While slightly more complicated than when there is X structure, eigenvalues can still be calculated and thereby the entropies without much difficulty.

In summary, the major part of this paper is to present a procedure for calculating quantum discord in an AB system when both sub-systems are of dimension larger than 2. A previously given recipe for two dimensions, as in Sec. 2, is generalized in Secs. 3 and 4 to higher $d$ with a focus on keeping to a minimum the number of parameters that need to be extremized, the four in Eq.~(\ref{eqn15}) with the two more in Eq.~(\ref{eqn16}) for $d=3$. Admittedly, our procedure is limited to classes of density matrices and not generally applicable in these higher dimensions with questions arising of POVMs and NP. Special attention is paid to the class of states called $X$-states and the second part of the paper is a study of their systematics in the Appendix below. 

\section*{Acknowledgments}

I thank the referees for substantive comments that improved this paper from its initial form. 
	
\section*{Appendix: Systematics of $X$-states}

For a $4 \times 4$ qubit-qubit system, $X$-states were introduced and named for their visual resemblance to the letter \cite{ref15}. But, as states using just 7 parameters out of the 15 of a general density matrix, they are characterized by their symmetry under the su(2) $\otimes$ u(1) $\otimes$ su(2) sub-algebra of the full su(4) algebra that applies to the pair of qubits, whether or not the density matrix looks like X \cite{ref16}. This perspective also permits the generalization to $X$-states of $n$ qubits. The recognition that for a single qubit, the $2 \times 2$ density matrix is immediately an $X$, and that the pair system involves two such $2 \times 2$ su(2) spaces with a phase between them, represented by the u(1), points to the same iteration when more qubits are added. Thus, for 3 qubits, where the full algbera is of su($2^3=8$) with $8^2-1=63$ parameters, the sub-algebra [su(2) $\otimes$ u(1) $\otimes$ su(2)] $\otimes$ [u(1)] $\otimes$ [su(2) $\otimes$ u(1) $\otimes$ su(2)], a 15-parameter sub-algebra of the full su(8) algebra gives their $X$-states, and so on \cite{ref18}. The number for $n$ qubits is $2^{n+1} -1$, the iteration described algebraically as the induction $2^{n+1}-1 = 2(2^n-1)+1$. This constitutes a sub-algebra of that dimension of the full su($2^n$) algebra of $n$ qubits. A connection to finite projective geometries \cite{ref36,ref37} is also worth noting.

Given the benefits of recognizing such symmetry structures and of sub-algebras to generate states for multiple qudits, we present here a similar development for qutrits or higher dimensional systems. As can be seen from the structure of Eq.~(\ref{eqn21}), a three-dimensional central $3 \times 3$ block of a single qutrit's $X$-state su$^{(X)}$(3) (with two real and one complex or four parameters) is embedded in six other dimensions around it. Thus, a qutrit-qutrit $X$-state has 16 parameters. As with the previous paragraph for qubits, the iteration sub-algebra of the full su(9) algebra (of 80 parameters) can be described as three copies now of the starting su$^{(X)}$(3) with u(1)s in between and at the ends for the 12+4 number of parameters of u(1) $\otimes$ [su$^{(X)}$(3)] $\otimes$ u(1) $\otimes$ [su$^{(X)}$(3)] $\otimes$ u(1) $\otimes$ [su$^{(X)}$(3)] $\otimes$ u(1). The next step similarly repeats thrice the previous with 4 additional u(1)s. For $n$ qutrits, the number of parameters is $2(3^n-1)$ and the inductive iteration is given by $2(3^n-1)=3[2(3^{n-1}-1)]+4$. Again, in the language of symmetry groups and algebras, these are sub-algebras of the full su($3^n$) algebra of $n$ qutrits.

Table II presents values for various classes of states of such pair systems of dimensions 2 and 3. Also shown for qubits are the number of su(2) and u(1) in the sub-algebra of $X$-states of $n$ qubits. Three times the former number plus the latter is, of course, the number shown in the row for $X$-states. That number is also the number of points in the projective geometry PG($n$, 2), whereas the number in the row above of the general state is PG($2n-1$, 2). In particular, PG(2, 2) with 7 points is the Fano Plane, the simplest projective plane, of extensive mathematical interest \cite{ref38}. It is interesting that the number of $X$-states of qubits runs through all successive integer values $n$ of PG($n$, 2) whereas the number for general states skips even values, depending as it does on $2n-1$. Alternatively, this can be stated in terms of skipping half-odd integer values of $n$, that is, that PG(2, 2), PG(4, 2), etc., are absent in the general count. There is a whiff of quantum-mechanical spin angular momentum in these sequences of numbers!

\begin{table}[h] 
\begin{center}
\begin{tabular}{|c|c|c|c|c|c|}
\hline
 number of qubits&1&2&3&4&n \\ 
 \hline
general&3&15&63&255&$2^{2n}-1$   \\
\hline
X-states&3&7&15&31&$2^{n+1}-1$   \\
\hline
su(2)s&1&2&4&8&$2^{n-1}$ \\
\hline
u(1)s&0&1&3&7&$2^{n-1}-1$ \\
\hline
extended X states&3&15&31&63&$2^{n+2}-1$ \\
\hline
\hline
number of qutrits&1&2&3&4&n \\
\hline
general&8&80&728&6560&$3^{2n}-1$ \\
\hline
X states&4&16&52&160&$2(3^n-1)$ \\
\hline
extended X states&4&44&152&476&$2(3^{n+1}-5)$ \\
\hline
\end{tabular}
\end{center}
\caption{Enumeration of the number of states in a general, an $X$-, and an extended $X$- state of $n$ qubit and qutrit systems.}
\end{table}

Other general enumeration of the number of parameters for arbitrary dimension follows from straightforward if slightly tedious algebra. As examples, a general qubit-qudit bipartite system has $4d^2-1$, $X$-state has $4d-1$, and extended $X$-state $8d-1$ for $d$ even and $8d-5$ for $d$ odd as the number of parameters. The results for a system of $n$ qubits follows by setting $d=2^{n-1}$ so that there are $2^{2n}-1$, $2^{n+1}-1$ and $2^{n+2}-1$ states for the general, $X$-, and extended $X$-, state, respectively. Similarly, a qutrit-qudit bipartite system has $9d^2-1$ for a general density matrix, $6d-1$ for $d$ even and $6d-2$ for $d$ odd in $X$-states, and $18d-1$ for $d$ even and $18d-10$ for $d$ odd in extended $X$-states. Again, setting $d=3^{n-1}$ gives the results for $n$ qutrits: $3^{2n}-1$, $2(3^n-1)$ and $2(3^{n+1}-5)$, respectively. 

Even more broadly, a qudit-qudit bipartite system has $d^4-1$ parameters in a general $\rho$, $2d^2-1$ for $d$ even and $2d^2-2$ for $d$ odd in $X$-states, and $2d^3-1$ for $d$ even and $2d^3-d^2-1$ for $d$ odd extended $X$-states. For $d=4$, this number for $X$-states of $n$ such qudits is $2(4^n)-1$ which means through $2(4^n)-1 = 4[2(4^{n-1})-1]+3$ that each successive qu4it means an iteration of 4 copies of the previous with u(1)s in between the copies, paralleling the iteration for $n$ qubits. Again, these have the symmetry of a sub-algebra of the large su($d^n$) algebra. An asymmetric qudit-quDit system has similar enumerations. There are in general $d^2D^2-1$ parameters, $X$-states having the smaller number $2dD-1$ for $dD$ even and $2dD-2$ for $dD$ odd. 

It is also interesting to note as a complement to the above paragraphs that for pure states, the density matrices have fewer parameters. With the eigenkets themselves available, a normalization and an overall phase dropping out, there are $d^n-1$ complex elements or twice that number of real parameters defining such a density matrix. This number is very similar to the number for corresponding $X$-states noted above. Finally, so-called quantum-classical density matrices are of interest in some discussions of correlations. With A described in terms of quantum ket-bra and B a classical density matrix, they have the form

\begin{equation}
\rho = \sum_{i=1}^d p_i |\psi_i\rangle \langle \psi_i| \rho_i^{(B)},
\label{eqn22}
\end{equation}
with $p_i$ probabilities ($0 \leq p_i \leq 1$), and $\rho^{(B)}$ classical density matrices of sub-system B. As noted for pure states, these are $2(d-1)$ in number, there are $d-1$ probabilities, and each classical density matrix has $D^2-1$ parameters for a total of $d(D^2+d-1)-1$ parameters, again smaller than the full $d^2D^2-1$ in a general $\rho$. Also, again the number is asymmetric in $d$ and $D$.

\end{document}